# Heralded Generation of High-Visibility Three-Photon NOON States

Heonoh Kim, Hee Su Park, and Sang-Kyung Choi

Korea Research Institute of Standards and Science (KRISS),

1 Doryong-dong, Daejeon 305-340, Republic of Korea

We describe an experimental demonstration of a novel three-photon NOON state generator using a single source of photons based on spontaneous parametric down-conversion (SPDC). The three-photon number-entangled state is deterministically generated by the detection of a heralding photon. Interference fringes measured with an emulated three-photon absorber reveal the three-photon de Broglie wavelength and exhibit visibility > 70% without background subtraction.

42.50.Dv, 42.50.St, 03.67.Bg, 42.50.Ex

Quantum optical technologies based on multi-photon entangled states can exceed the fundamental limits such as the Rayleigh diffraction limit of optical imaging [1] or the standard quantum limit of optical phase estimation [2-5]. In particular, the resolution of lithography and metrology can be improved beyond the classical limit by applying what is known as a NOON state, described as $|\Psi\rangle = 1/\sqrt{2}(|N,0\rangle + |0,N\rangle)$, which embodies photon-number entanglement between two distinct optical modes [1-10]. These $N$ photons comprising a NOON state can collectively behave as a single photon whose wavelength is $1/N$ of the single-photon wavelength, thereby effectively increasing the imaging resolution and the phase estimation uncertainty by a factor of $1/N$ and $1/N^{1/2}$, respectively [1-5]. A NOON state generator is also applicable to the generation of few-photon polarization-squeezed light that can be 'over-squeezed' [11]. The simplest NOON state with $N=2$ has been experimentally demonstrated since 1990 utilizing an SPDC photon pair and a Hong-Ou-Mandel (HOM) interferometer [6-8,12]. The potential benefits of an *N*=2 state are, however, limited by the relevant de Broglie wavelength [13] being equal to, hence no shorter than the SPDC pump wavelength. Various theoretical schemes for generating NOON states with *N*>2 have been proposed [14-21], but because of technical difficulties [22,23] only a handful of experiments have so far realized *N*>2 NOON states [5,9,10,24-26]. Moreover, most of these experiments utilize state projection measurements to suppress the effect of superfluous states. The state projection method enables super-resolution interferometry even with classical light sources [26]; however, it is incompatible with the original quantum lithography proposal that requires *N*-photon detection in a single spatial mode [1]. To the best of the authors' knowledge, only one work in the literature reports super-resolution interference measurements using an *N*>2 NOON state without relying on state projection [9].

The major technical difficulty for experimental realization of NOON states is the lack of practical near-ideal photon sources. Theoretical proposals typically assume the deterministic generation of a definite number of photons [14-19]. However, practical sources based on SPDC or faint laser pulses generate photons in probabilistic fashion, which inevitably produces redundant photons that degrade the visibility of NOON state interference fringes [9]. Therefore, minimizing the number of independent probabilistic photon sources potentially improves the quality of super-resolution interference. The first demonstration of the three-photon NOON state used two sources, an SPDC source and a weak laser pulse [9]. The super-resolution quality can be further improved by generating the photons from a single probabilistic source. This Letter reports the experimental realization of a novel *N*=3 NOON state generation scheme based on double photon pair emission from a single SPDC source.

A schematic of our NOON state generator is shown in Fig. 1. Pulse-pumped non-collinear type-I SPDC produces two indistinguishable pairs of horizontally polarized photons. We note that the polarization beam splitters (PBSs) transmit (reflect) horizontally (vertically) polarized photons, and wave plate angles given henceforth denote the inclination of the slow axis with respect to the horizontal axis. The photons along the upper path have their polarization changed to the vertical a combination of a mirror and a quarter-wave plate (QWP1) aligned along 45°, and merge with the two lower-path photons at PBS2 into a single spatial mode. After these photons pass through a half-wave plate (HWP1) with angle at 22.5°, the four-photon state becomes

$$|\Psi\rangle_{\text{HWP1}} = \left(\frac{1}{8}a_H^{\dagger 4} - \frac{1}{4}a_H^{\dagger 2}a_V^{\dagger 2} + \frac{1}{8}a_V^{\dagger 4}\right)|0\rangle, \quad (1)$$

where $a_H$ ($a_V$) is the annihilation operator for the horizontal (vertical) polarization mode. Horizontally polarized photons are wholly transmitted by the partial PBS (PPBS), while 1/3 (2/3) of the vertically polarized photons are transmitted (reflected). When exactly one photon is reflected by the PPBS, the remaining three photons propagating to QWP2 are in the state

$$|\Psi\rangle_{\text{PPBS}} = \left(-\frac{\sqrt{2}}{6}a_H^{\dagger 2}a_V^{\dagger} + \frac{\sqrt{2}}{18}e^{2i\phi}a_V^{\dagger 3}\right)|0\rangle, \quad (2)$$

where $\phi$ is the magnitude of the birefringence introduced by PPBS between the horizontal and vertical polarizations. Adjusting the angles of QWP2 and HWP2 to 45° and $\theta = \phi/4$, respectively, transforms the output from HWP2 into the three-photon NOON state described as

$$|\Psi\rangle_{\text{HWP2}} = \frac{1}{9}\left(ia_H^{\dagger 3} + a_V^{\dagger 3}\right)|0\rangle = \frac{\sqrt{6}}{9}\left(i|3_H,0_V\rangle + |0_H,3_V\rangle\right). \quad (3)$$

Barring losses, the probability for the initial double photon pair to yield the above state is 4/27, and the remaining cases produce states that send zero or multiple photons to the heralding single photon counter 1 (SPC1).

Let us estimate the robustness of our $N=3$ NOON state generation scheme against performance degradation by redundant photons. A useful figure of merit is the ratio $P(N_{ex}=0)/P(N_{ex}>0)$, where $P(N_{ex})$ is the probability per pump pulse to produce an adequate number of photons with $N_{ex}$ photons in excess of requirement. If $\gamma$ is the probability for a pump pulse to generate an SPDC photon pair, then the double pair and triple pair generation probabilities are $\gamma^2$ and $\gamma^3$, respectively, and $P(N_{ex}=0)/P(N_{ex}>0) \approx 1/\gamma = P(N_{ex}=0)^{-1/2}$. This compares with $P(N_{ex}=0)/P(N_{ex}>0) \approx (\gamma/\alpha + \alpha/2)^{-1} < P(N_{ex}=0)^{-1/3}$ for a scheme based on an SPDC photon pair plus one photon generated with probability α from a weak coherent pulse [9]. Therefore, our scheme based on double SPDC photon pair emission allows a higher NOON state generation probability for a given $P(N_{ex}=0)/P(N_{ex}>0)$ ratio.

The SPDC source in Fig. 1 consists of a type-I phase-matched BBO crystal (thickness 1 mm) pumped by a periodic UV pulse train (center wavelength 390 nm; pulse duration 200 fs; repetition rate 76 MHz; average power 500 mW) from a frequency-doubled mode-locked Ti:sapphire laser. The pump beam is focused onto a spot with a full-width at half-maximum (FWHM) diameter of 50 μm in the BBO crystal, and the down-converted photons are subsequently collimated with convex lenses with a focal length of 150 mm. The PPBS is a cube-type PBS made of custom-coated BK7 glass whose measured transmissions are 99% and 31% for horizontal and vertical polarization inputs, respectively. An interference filter (center wavelength 780 nm; FWHM bandwidth 5 nm) is inserted between PBS2 and the PPBS to reduce background noise and improve the degree of temporal/spectral indistinguishability among the four SPDC photons.

The heralding photon is coupled into a single-mode fiber (SMF: mode field diameter 5.6 μm ; numerical aperture 0.12) and counted with a single photon counter (SPC1: Perkin-Elmer, SPCM-AQ4C). The interference between the horizontal and vertical polarization components of the output state from HWP2 is measured by means of QWP3, HWP3, PBS3, and two SMF directional couplers connected to three SMF-coupled SPCs (SPC2~SPC4). QWP3 set at 45° converts the horizontal and vertical polarizations into the left- and right-circular polarizations, respectively, while HWP3 swaps these two circular polarizations and applies an angle-dependent phase difference between them. Horizontally polarized photons due to interference between the two circular-polarization components are transmitted by PBS3 and coupled into the SMF using an aspheric lens (focal length 8.0 mm). Both SMF directional couplers have a beam-splitting ratio of 43:57, hence the three photons entering the SMF divide equally among the three SPCs with a probability of 21%. The coincidence counting by these three SPCs effectively emulates three-photon absorption in a single spatial mode.

The preparation of the NOON state is initialized by adjusting the position of the back-reflecting mirror with HOM interferometry. For this purpose only, the average pump power for SPDC is lowered to 100 mW to generate mostly single pairs of photons. Coincidence counting of SPC1 and SPC2 with the angles of QWP2, HWP2, QWP3, and HWP3 set uniformly along 0° projects the state entering PPBS to $|1_H,1_V\rangle$. Spatio-temporal mode overlap between the upper and lower path photons is achieved by obtaining an HOM dip with a visibility of 0.97, as shown in Fig. 2(a). The HWP2 angle is tuned to maximize four-photon coincidence counts with QWP2, QWP3, and HWP3 aligned along 45°, 0°, and 0°, respectively. These coincidence counts are proportional to the probability that all three photons are horizontally polarized before entering QWP3. As shown in Fig. 2(b), the optimal HWP2 angle is a multiple of 45°, which implies that the PPBS does not introduce significant birefringence.

The interference between the horizontal and vertical polarization modes measured with respect to the prepared $N$=3 NOON state are shown in Fig. 3. The QWP3 angle has been reset to 45°, and the interference fringes are obtained by varying the HWP angle, which equals one-fourth the phase difference between the two principal polarization components. Figure 3(a) shows the two-fold coincidence counts of SPC1 and SPC2. The interference fringes in Fig. 3(a) are due to single-photon interference between the horizontal and vertical polarization components of the right-circularly polarized photons entering QWP3. Figure 3(b) shows four-fold coincidence counts of SPC1~SPC4, which reveal super-resolution interference with fringes that oscillate three times faster compared to single-photon interference. The solid line is a sinusoidal fit to the experimental data, and the error bars represent the standard deviations that correspond to the square roots of the measured counts. The visibility of the three-photon interference fringes is 0.72±0.03 without background subtraction, which clearly surpasses the level of visibility (10% for $N$=3) achievable with a proposed technique based on classical nonlinear optics [27]. The unheralded three-fold coincidence counts of SPC2~SPC4 shown in Fig. 3(c) are, on average, two orders greater than in Fig. 3(b), and mostly attributable to the double pair of SPDC photons transmitting through the PPBS with no photon going to SPC1. In this case, all four photons incident on HWP3 are horizontally polarized and produce a sharp peak when the phase difference between the principal polarization components is a multiple of 360°. The single photon detection probability at SPC1 is measured to be $(1.92\pm0.02)\times10^{-3}$, and this multiplied by the unheralded three-photon coincidence probability approximately equals the four-photon coincidence probability due to triple pair production by the SPDC source. Subtracting our estimated triple-pair background from the interference pattern of Fig. 3(b) yields Fig. 3(d), which exhibit three-photon interference with a visibility of 0.91±0.03. The three-photon interference visibility we have obtained indicates that the fidelity $\langle\Psi_{ideal}|\rho_{exp}|\Psi_{ideal}\rangle$ between the generated $\rho_{exp}$ and the ideal NOON state $|\Psi_{ideal}\rangle$ is greater than 0.68±0.03 (0.90±0.03 with background subtraction) [28].

In conclusion, we have demonstrated the generation of heralded three-photon NOON states using a novel scheme based on linear optics and a single SPDC source that emits two pairs of photons. Three-photon number entanglement between two orthogonal polarization modes has been verified by observing interference fringes with a de Broglie wavelength three times smaller than the single-photon wavelength. The level of interference visibility, even without the aid of state projection or background subtraction, exceeds the present expectation of classical nonlinear optics. Our scheme can be extended to generate $N>3$ NOON states. For example, an $N=4$ NOON state can be generated from a triple photon-pair state $|3_H,3_V\rangle$ by replacing the PPBS in our scheme with two PPBSs having reflectances of 1/3 and 0 for the vertical and horizontal polarizations, respectively, and one HWP at 45° in between. Simultaneous detection of the single photons reflected by each PPBS heralds the generation of an $N=4$ NOON state after HWP2 with a probability of 16/243 for a given $|3_H,3_V\rangle$. The structure of this $N=4$ NOON state generation scheme turns out to be topologically equivalent to the proposal by Lee *et al*. [14]. The method we have outlined is essentially a procedure for subtracting heralding photons from an initial product state so that the remaining photons are projected to a desired quantum state [29]. Although the generation efficiency decreases as $N$ increases, our extendable scheme is currently an experimentally feasible way to realize higher NOON states and other kinds of nonclassical states as well.

This work was supported by the KRISS project 'Single-Quantum-Based Metrology in Nanoscale.' H. K. and H. S. P. contributed equally to this work.

hspark@kriss.re.kr

[1]  A. N. Boto *et al.*, Phys. Rev. Lett. **85**, 2733 (2000); P. Kok *et al.*, Phys. Rev. A **63**, 063407 (2001).

[2]  Z. Y. Ou, Phys. Rev. A **55**, 2598 (1997).

[3]  K. Edamatsu, R. Shimizu, and T. Itoh, Phys. Rev. Lett. **89**, 213601 (2002).

[4]  D. Leibfried *et al.*, Science **304**, 1476 (2004).

[5]  T. Nagata *et al.*, Science **316**, 726 (2007).

[6]  J. G. Rarity *et al.*, Phys. Rev. Lett. **65**, 1348 (1990).

[7]  M. D'Angelo, M. V. Chekhova, and Y. H. Shih, Phys. Rev. Lett. **87**, 013602 (2001).

[8]  Y. Kawabe *et al.*, Opt. Express **15**, 14244 (2007).

[9]  M. W. Mitchell, J. S. Lundeen, and A. M. Steinberg, Nature **429**, 161 (2004).

[10] P. Walther *et al.*, Nature **429**, 158 (2004).

[11] L. K. Shalm, R. B. A. Adamson, and A. M. Steinberg, Nature **457**, 67 (2009).

[12] C. K. Hong, Z. Y. Ou, and L. Mandel, Phys. Rev. Lett. **59**, 2044 (1987).

[13] J. Jacobson, G. Björk, I. Chuang, and Y. Yamamoto, Phys. Rev. Lett. **74**, 4835 (1995).

[14] H. Lee, P. Kok, N. J. Cerf, and J. P. Dowling, Phys. Rev. A **65**, 030101(R) (2002); P. Kok, H. Lee, and J. P. Dowling, Phys. Rev. A **65**, 052104 (2002).

[15] J. Fiurasek, Phys. Rev. A **65**, 053818 (2002).

[16] G. J. Pryde and A. G. White, Phys. Rev. A **68**, 052315 (2003).

[17] H. F. Hofmann, Phys. Rev. A **70**, 023812 (2004).

[18] N. M. VanMeter *et al.*, Phys. Rev. A **76**, 063808 (2007).

[19] H. Cable and J. P. Dowling, Phys. Rev. Lett. **99**, 163604 (2007).

[20] A. E. B. Nielsen and K. Mølmer, Phys. Rev. A **75**, 063803 (2007).

[21] H. F. Hofmann and T. Ono, Phys. Rev. A **76**, 031806(R) (2007).

[22] B. L. Higgins *et al.*, Nature **450**, 393 (2007).

[23] A. Cho, Science **312**, 672 (2006).

[24] F. W. Sun, Z. Y. Ou, and G. C. Guo, Phys. Rev. A **73**, 023808 (2006); F. W. Sun, B. H. Liu, Y. F. Huang, Z. Y. Ou, and G. C. Guo, Phys. Rev. A **74**, 033812 (2006).

[25] B. H. Liu *et al.*, Phys. Rev. A **77**, 023815 (2008).

[26] K. J. Resch *et al.*, Phys. Rev. Lett. **98**, 223601 (2007).

[27] S. J. Bentley and R. W. Boyd, Opt. Express **12**, 5735 (2004).

[28] The visibility determines the lower bound of the magnitude of the off-diagonal density matrix element, and therefore leads to the lower bound of the fidelity.

[29] M. Dakna, T. Anhut, T. Opatrny, L. Knoll, and D. G. Welsch, Phys. Rev. A **55**, 3184 (1997).

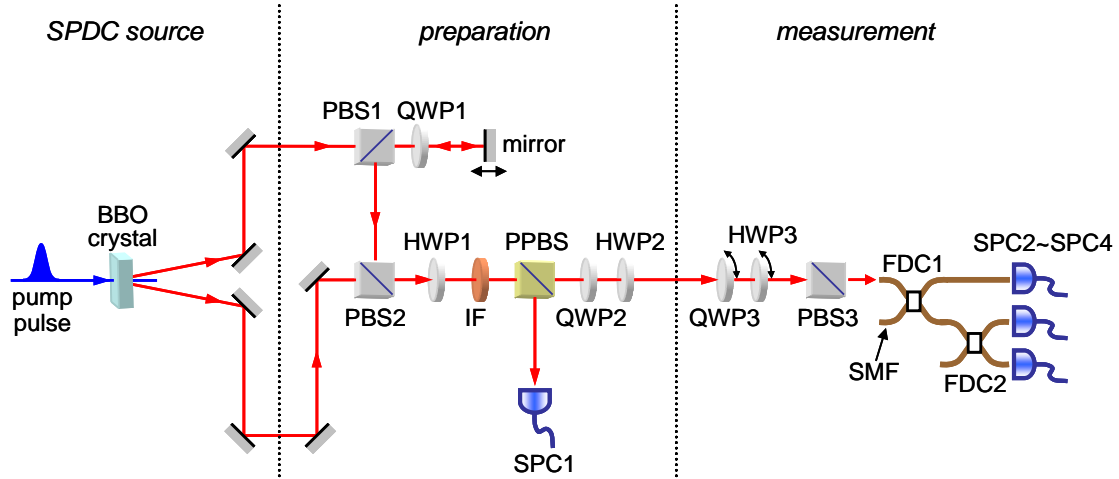

FIG. 1 (color online). Generation and measurement of heralded three-photon NOON states. PBS: polarizing beam splitter; PPBS: partial PBS; HWP: half-wave plate; QWP: quarter-wave plate; IF: interference filter; SPC: single photon counter; SMF: single-mode fiber; FDC: fiber directional coupler.

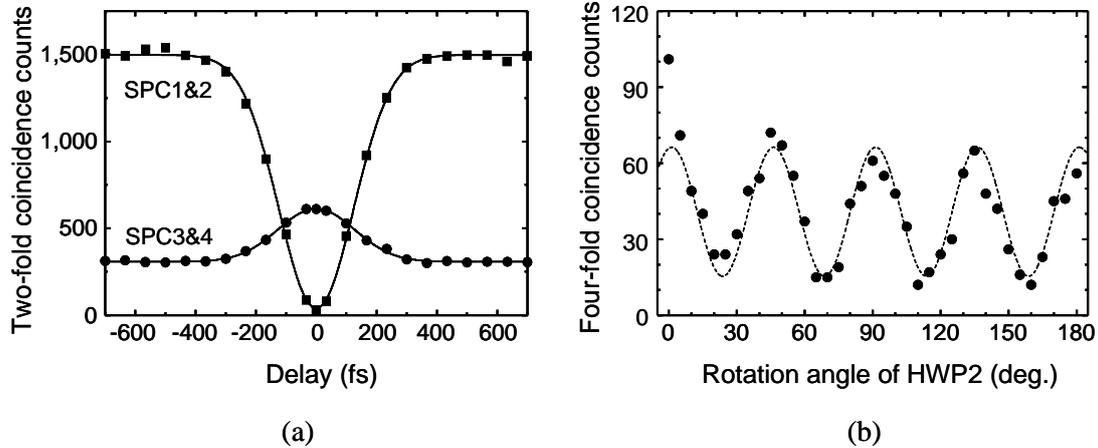

FIG. 2. Preparation of the three-photon NOON states: (a) HOM interferometry to match the two optical path lengths: two-fold coincidence counts per 1 s as a function of the time delay between the photons entering PBS2 through the upper and the lower paths. The solid curves are Gaussian least-square fits. (b) Adjustment of the HWP2 angle: four-fold coincidence counts of SPC1~SPC4 per 300 s. The angles of QWP2, QWP3, and HWP3 are set at 45°, 0°, and 0°, respectively. The dashed line is a sinusoidal fit.

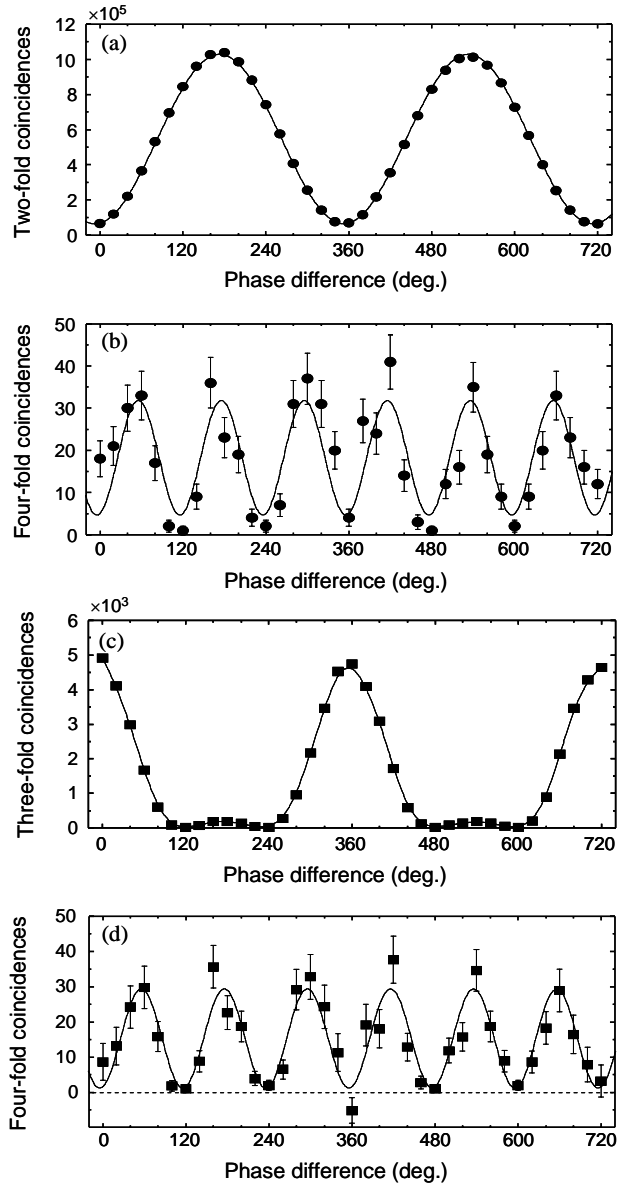

FIG. 3. Measurement results of single- and three-photon interference. The data points are coincidence counts per 300 s, and the error bars represent standard deviations calculated as $(counts)^{1/2}$. (a) Single-photon interference: two-fold coincidence counts of SPC1 and SPC2; (b) three-photon interference by the heralded states: four-fold coincidence counts of SPC1~SPC4; (c) three-photon interference by the non-heralded states: three-fold coincidence counts of SPC2~SPC4; (d) three-photon interference by the heralded states after subtracting the background coincidences due to the triple-pair generation by SPDC.